# Letter: Mathematical Foundation of Turbulence Generation−Symmetric to Asymmetric Liutex/Rortex *

Jian-ming Liu[1,2], Yue Deng[3], Yi-sheng Gao[2], Sita Charkrit[2], Chaoqun Liu[2*]
1. School of Mathematics and Statistics, Jiangsu Normal University, Xuzhou 221116, China
2. Department of Mathematics, University of Texas at Arlington, Arlington 76019, USA
3. Department of Physics, University of Texas at Arlington, Arlington 76019, USA

**Abstract:** Vortex has been considered as the building block and muscle of turbulence for long time. A new physical quantity called Liutex (previously named Rortex) has been defined as the rigid rotation part of fluid motion [C. Liu *et al.*, "Rortex—A new vortex vector definition and vorticity tensor and vector decompositions", Phys. Fluids 30, 035103 (2018)]. Since turbulence is closely related to vortex/rotation, it is concluded that there is no turbulence without Liutex and there is no turbulence without asymmetric Liutex. From DNS and experiment, forest of hairpin vortices has been found in flow transition and low Reynolds turbulence, but more and more one leg vortices appear in fully developed turbulence. This letter shows hairpin vortex is unstable. Hairpin vortex will weaken or lose one leg by the shear and vortex interaction. This conclusion is made by Liutex definition and mathematical analysis without any physical assumptions. The asymmetry of vortex is caused by the interaction of symmetric shear and symmetric Liutex due to the definition of Liutex, which considers the smaller element of a pair of vorticity elements as the rotational strength. For a 2-D fluid rotation, if a disturbance shear effects the larger element, the rotation strength will not be changed, but if the disturbance shear effects the smaller element, the rotation strength will be immediate changed due to the Liutex strength definition. For a rigid rotation, if both vorticity of the shear and Liutex have the same directions, e.g., clockwise, the Liutex strength will not be changed. If the vorticity of the shear and Liutex have different directions, e.g., one clockwise and one counterclockwise, the Liutex strength will be weakened. Then, the hairpin vortex could lose the symmetry and even deform to a one leg vortex. The one leg vortex cannot keep balanced, and the chaotic motion and flow fluctuation are doomed. This is considered as the mathematical foundation of turbulence formation. This theory has been checked and proved by our DNS results of boundary layer transition.

**Introduction**
Turbulence is one of the most complex phenomena in nature. It is widely found in the atmosphere, ocean engineering, aviation, aerospace, and industrial engineering [1,2]. In 1883, Reynolds revealed this complex flow pattern through his famous round tube experiment. For the next over one hundred years, a large number of scientists engaged in turbulence research and solved a large number of engineering problems. However, due to the complexity of turbulence, people have not been able to reveal the universal mechanism of turbulence generation, especially about asymmetry which is a crucial step of turbulence formation. This leads to various turbulence generation theories. Richardson gave a classical turbulence theory by vortex chains generated by large vortex breakdown [3]. But from the direct numerical simulations (DNS), the vortex chain is never observed. Kolmogorov accepted the Richardson energy cascade and vortex breakdown and thought that the large eddies passed energy to small eddies through vortex breakdown, then continue to smaller, until to reach the Kolmogorov's smallest scale [4]. But by using of the most accurate experimental equipment until now, one cannot find turbulence is caused by the vortex breakdown. Up to now, no effective mathematical principle has been found to explain the generation of turbulence, especially the mathematical principle for symmetrical vortices to become asymmetric in the boundary layer flow transition process. This letter attempts to give a mathematical principle of turbulence generation based on the bias definition of the Liutex/Rortex vector.

---

 * Project supported by the National Nature Science Foundation of China (Grant No. 91530325).
**Biography:** Jian-ming Liu (1977-), Male, Ph. D., Associate Professor, E-mail: jmliu@jsnu.edu.cn
**Corresponding author:** Chaoqun Liu, E-mail: cliu@uta.edu

## 1. Necessary condition for turbulence

Turbulence or turbulent flow is a common type of fluid motion characterized by chaotic changes in pressure and flow velocity. Turbulence is commonly observed in everyday phenomena and most realistic engineering flows [1-2]. Richard Feynman has described turbulence as the most important unsolved problem in classical physics [3]. Turbulent flow is very irregular, diffusive, dissipative and chaotic. Vortex is the building block and muscle of turbulence with variety of vortices of different sizes and rotational strengths (Liu et al, 2014) [4]. Without vortex, there would be no turbulence. Without asymmetric vortex, there would be no turbulence.

## 2. Definition of Liutex $\vec{R}$

Many vortex identification methods were developed during the past decades. Among them, the new Omega vortex identification method could be one of the best [4-9]. Recently, a new physical quantity called Liutex has been given to represent the rigid rotation of fluid motion (Liu et al, 2018; Gao et al, 2018) [10,11]. Liutex $\vec{R}$ is a vector that is uniquely defined by $\vec{R} = R\vec{r}$. Its direction is defined by the real eigenvector of velocity gradient tensor $\nabla \vec{u}$ and its magnitude $R$ is defined as the angular speed of rigid rotation, i.e.,

$$R = \langle \vec{\omega}, \vec{r} \rangle - \sqrt{\langle \vec{\omega}, \vec{r} \rangle^2 - 4\lambda_{ci}^2} \quad \text{and} \quad \nabla \vec{u} \cdot \vec{r} = \lambda_r \vec{r}$$

where $\vec{r}$ is the direction of Liutex vector [12,13], $\vec{\omega}$ is vorticity and $\lambda_{ci}$ is the imaginary part of the complex eigenvalue of $\nabla \vec{u}$ and $\vec{\omega} \cdot \vec{r} > 0$. From our previous work, the RS decomposition was presented by

$$\nabla \vec{V}_{\theta\min} = \begin{bmatrix} \lambda_{cr} & -\phi & 0 \\ \phi + s & \lambda_{cr} & 0 \\ \xi & \eta & \lambda_r \end{bmatrix} = \overline{\overline{R}} + \overline{\overline{S}} \quad (1)$$

$$\overline{\overline{R}} = \begin{bmatrix} 0 & -\phi & 0 \\ \phi & 0 & 0 \\ 0 & 0 & 0 \end{bmatrix} \quad (2)$$

$$\overline{\overline{S}} = \begin{bmatrix} \lambda_{cr} & 0 & 0 \\ s & \lambda_{cr} & 0 \\ \xi & \eta & \lambda_r \end{bmatrix} = \begin{bmatrix} 0 & 0 & 0 \\ s & 0 & 0 \\ \xi & \eta & 0 \end{bmatrix} + \begin{bmatrix} \lambda_{cr} & 0 & 0 \\ 0 & \lambda_{cr} & 0 \\ 0 & 0 & \lambda_r \end{bmatrix} \quad (3)$$

we picked the magnitude of $\vec{R}$ as the $2\min\{|-\phi|,|\phi+s|\}$ and $R = 2\phi$, assuming both $\phi$ and $s$ are positive.

## 3. Justification of definition for magnitude of Liutex

A correct definition must stand for both 2-D and 3-D cases. Let us take a 2-D laminar channel flow solution (Fig. 1a) as an example to justify our definition of the magnitude of Liutex. Then, the exact solution is $u = (1-y)(1+y), v = 0$ and then $\partial u/\partial x = 0$, $\partial u/\partial y = -2y$, $\partial v/\partial x = 0$, and $\partial v/\partial y = 0$. Thus we can get the vorticity $\omega_z = 2y$. In order to define the magnitude of Liutex/Rortex which should measure the strength of the rigid rotation or angular speed, we have several choices like the maximum, the minimum, or the average. Since there is no rotation, it is apparently inappropriate to pick the maximum or the average which is really a vorticity component $\omega_z$. According to the definition of Liutex, $R = 2\min\{0,|-2y|\} = 0$, which is appropriate to describe the rotation strength. Since there is no rotation in the laminar channel flow. The boundary layer solution on a flat plate or Blasius solution (Fig. 1b) will lead same conclusion that Liutex magnitude should be $R = 2\min\{|\partial u/\partial y|,|\partial v/\partial x|\} = 0$. That is there is no rotation in a laminar boundary layer. Furthermore, if a rigid rotation (Fig. 1c) is considered, we have $u = \omega y$ and $v = -\omega x$, $R = 2\omega$, $\omega$ is the exact angular speed. As a result, the shear can be presented by $\tau_{xy} = (\partial u/\partial y + \partial v/\partial x)/2 = 0$ and there is no energy dissipation in the rigid rotation.

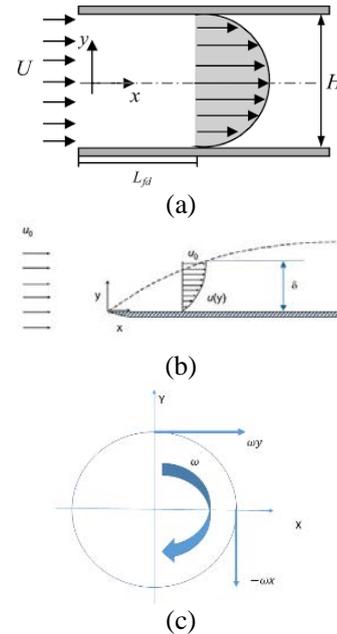

Fig. 1 (a) 2-D channel flow, (b) boundary layer flow and (c) Rigid rotation

## 4. Interaction of 2D shear and Liutex

A pure shear tensor can be described as $\begin{bmatrix} 0 & s \\ 0 & 0 \end{bmatrix}$ or $\begin{bmatrix} 0 & 0 \\ s & 0 \end{bmatrix}$ for 2-D. Let us add the disturbance shear to the channel flow, we will get $\begin{bmatrix} 0 & -2y+s \\ 0 & 0 \end{bmatrix}$ or $\begin{bmatrix} 0 & -2y \\ s & 0 \end{bmatrix}$. Although $R$ is zero at the beginning, the first shear does not generate Liutex, but the second one does. This clearly shows shear may or may not generate rotation depending on the shear direction. If both disturbance shear stresses increase the larger element of the base shear/Liutex, there is no change in the rotation strength. If the disturbance shear decreases the smaller element of the base shear/Liutex, the rotation strength will be reduced. Let us consider the interaction of a shear and a rigid rotation in which the tensor of the rigid rotation is $\begin{bmatrix} 0 & \omega \\ -\omega & 0 \end{bmatrix}$ and the shear is $\begin{bmatrix} 0 & s \\ 0 & 0 \end{bmatrix}$. By adding them together, we will have $\begin{bmatrix} 0 & \omega+s \\ -\omega & 0 \end{bmatrix}$. Assume $\omega$ is positive, we will have two totally different cases: (A) If $s$ is positive, $R$ will have no change no matter how big $s$ is, according to the definition of $R = 2\min\{|\partial u/\partial y|, |\partial v/\partial x|\}$; (B) If $s$ is negative, we will have several possibilities: (1) If $|s|<\omega$, then $R = 2\{\omega-|s|\}$; (2) If $|s|=\omega$, then $R = 2\{\omega-|s|\} = 0$; (3) If $|s|>\omega$, then $R = 0$ since $(-\omega)(\omega+s) > 0$. From this example, we can clearly find that the interaction of shear and Liutex could make no changes on $R$ if disturbance shear has the same direction as the rotation; however, $R$ will be reduced or even the fluid rotation is stopped if the shear has the opposite direction from the rotation. Now, consider another example with $\begin{bmatrix} 0 & \omega+s_1 \\ -\omega & 0 \end{bmatrix} + \begin{bmatrix} 0 & s_2 \\ 0 & 0 \end{bmatrix}$. Assume $\omega$ and $s_1$ are positive: (1) If $s_2$ is positive, $R = 2\omega$ will have no change; (2) If $s_2 < 0$ and $|s_2|<|s_1|$, then $R = 2\omega$ has no change either; (3) If $s_2 < 0$, $|s_2|>|s_1|$ and $\omega + s_2 + s_1 > 0$, then $R = 2(\omega + s_2 + s_1)$, otherwise $R = 0$. From these examples, we can find several questions and answers: (1) Does shear change Liutex? Not sure and it depends; (2) If shear has the same sign as the rotation, the rotation strength will not change; (3) If the shear and rotation have opposite signs with $|s_2|<|s_1|$, the rotation strength will still not change; (4) If the shear and rotation have opposite signs with $|s_2|>|s_1|$, however, the rotation strength will change. This gives a clue that the interaction of shear and Liutex is not reversible: the increase of shear does not increase the strength of rotation if the larger element increases, but the decrease of shear may reduce the strength of rotation. In any cases, the change of smaller element by shear will cause the change of the rotation strength. Although these are analyses for 2-D cases, this basic physics of shear and Liutex interaction should be similar to 3-D cases.

**5. From hairpin vortex to one leg vortex**

Let us take a look at interaction of shear and Liutex in 3-D flows. The tensor formula could be:

$$\overline{\overline{R}} = \begin{bmatrix} 0 & -\phi & 0 \\ \phi & 0 & 0 \\ 0 & 0 & 0 \end{bmatrix} \quad \overline{\overline{S}} = \begin{bmatrix} 0 & 0 & 0 \\ s & 0 & 0 \\ 0 & 0 & 0 \end{bmatrix},$$

and

$$\overline{\overline{R}} + \overline{\overline{S}} = \begin{bmatrix} 0 & -\phi & 0 \\ \phi+s & 0 & 0 \\ 0 & 0 & 0 \end{bmatrix}$$

The conclusion should be the same as in 2-D case. If $s$ is positive (the shear and rotation have the same directions), the magnitude of Liutex $R$ will not change at all, which means the shear cannot change the strength of rotation. On the other hand, if $s$ is negative (the shear and rotation have the opposite directions), the magnitude of Liutex or the strength of rotation will be weakened or even disappeared.

For the interaction of shear and non-rigid rotational vortex, which is very common in a boundary layer, we should have

$$\overline{\overline{R}} + \overline{\overline{S}}_1 + \overline{\overline{S}}_2 = \begin{bmatrix} 0 & -\phi & 0 \\ \phi+s_1+s_2 & 0 & 0 \\ 0 & 0 & 0 \end{bmatrix}$$

and assume both $\phi$ and $s_1$ are positive. The same conclusion will be achieved just like what we discussed above for the 2-D case.

Let us consider the interaction of a shear and a hairpin vortex. Both are symmetric. Assume the direction of shear is clockwise which we think $s$ is positive, and the hairpin has two counter-rotating legs with the right leg clockwise and the left leg counterclockwise. Interacted with a clockwise shear, the right leg will keep the same rotation strength or $R$, but the left leg may be weakened (becomes thinner) or even disappear. This process will make the original symmetric hairpin becomes an asymmetric hairpin or even a one-leg vortex. One example with the symmetric Liutex is shown in Fig. 2. Moreover, Fig. 3 shows the one-leg vortex ring appearing in the up-level of the boundary layer and Fig. 4 depicts a secondary vortex ring with one strong leg and one weak leg in the lower boundary layer.

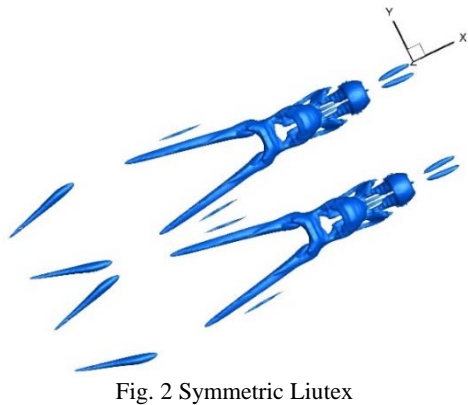

Fig. 2 Symmetric Liutex

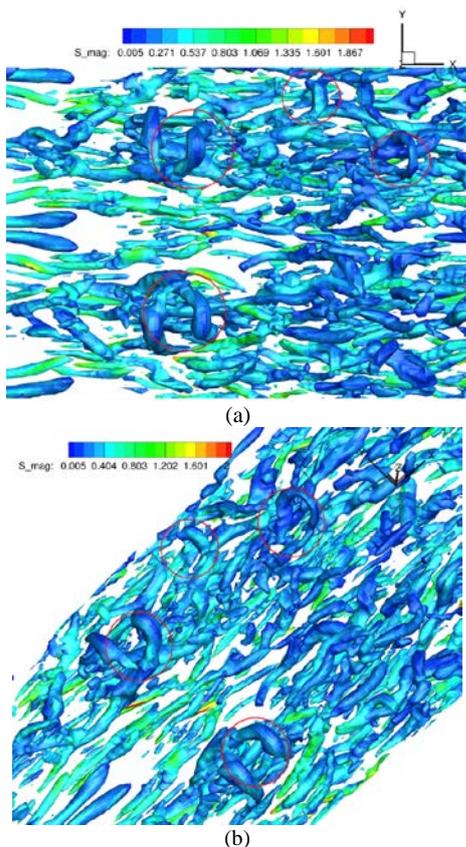

Fig. 3 One leg vortex or asymmetric vortices in the upper boundary layer (Liutex iso-surfaces colored by shear magnitude)

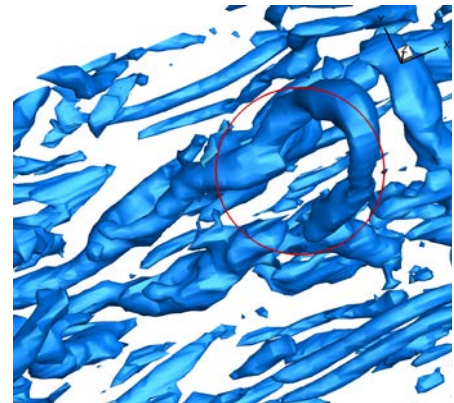

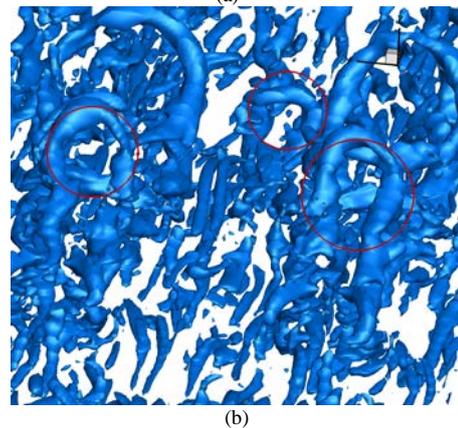

Fig. 4 One leg vortex in the second level vortex rings near the wall (Liutex iso-surfaces)

Confirmed by DNS and experiments, there are forest of hairpin vortices in the flow transition and early turbulence, but the hairpin vortex could be deformed or degenerated in the lower boundary layer where the viscosity is large or in fully developed turbulence zones due to the shear interaction with legs (Fig. 3). Note that the condition for the deformation or degeneration of symmetric hairpin is the existence of symmetric shears. That is the reason why in the inviscid flow region, the hairpin vortex keeps symmetric for a long time but the hairpin vortex in the lower boundary layer could quickly lose one leg. The only condition is the existence of fluctuated shear. If the shear moves in a clockwise motion, the clockwise vortex leg will not be affected. However, the counterclockwise vortex leg will be weakened or even disappeared. The asymmetric shear-Liutex interaction will cause the asymmetries of the hairpin vortices and further generate more to two asymmetric legs with one strong and one weak or even a one leg vortex. These could happen on the top of hairpin vortices (see Fig. 3) or secondary vortices located in the lower boundary layers (see Fig. 4).

## 5. Asymmetric Liutex and Chaotic

As confirmed by both DNS and experiment, there are many one leg vortices inside the lower boundary layer and fully turbulence. The one leg vortex cannot keep static as the nature of Liutex, which keeps rigid rotation. The asymmetric Liutex will keep unbalanced and keep swinging, which will produce asymmetry with fluctuating, swinging, shaking and chaos. As we addressed early, there is no turbulence if we have no Liutex and no asymmetric Liutex. However, shear is always in the boundary layer, especially in the lower boundary layer and hairpin vortices always appear in the flow transition and early turbulence (see Fig. 5). Unfortunately, the interaction of hairpin vortex and shear will cause non-symmetry due to the nature of shear and vortex interaction. Therefore, asymmetry, the one leg vortex, shaking of asymmetric vortices, and chaos are doomed. In other words, turbulence is doomed and that is the nature.

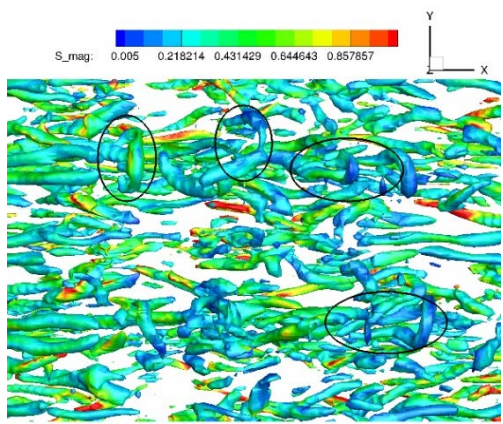

Fig. 5 Liutex iso-surfaces colored by shear magnitude

**6. Conclusions**

According to the description above, the following conclusion can be achieved as follows: (1) Liutex can represent a rigid fluid rotation; (2) The strength of Liutex is taken as a minimum and in a 2-D plane, described by $R = 2\min\{|\partial u/\partial y|, |\partial v/\partial x|\}$, (3)

Shear will not change the rotation strength $R$ if shear and $\vec{R}$ have the same directions, but shear may reduce $R$ if they have opposite directions and the shear magnitude is larger than the original shear contained in the original vortex; (4) The symmetric hairpin vortex may lose its symmetry when it interacts with symmetric shear; (5) A one leg hairpin vortex can be weakened or disappear due to the shear-vortex interaction. Therefore, a hairpin vortex is unstable in a boundary layer; (6) One leg or asymmetric vortices are shaking, swinging, chaotic, and then cause turbulence; (7) The nature of Liutex magnitude definition (smaller element of a pair) and interaction of shear and vortex is the mathematical foundation of turbulence generation; therefore, the symmetry loss and chaos are doomed.

**Acknowledgments**

This work was mainly supported by the Department of Mathematics of University of Texas at Arlington. The authors are grateful to Dr. Liandi Zhou for beneficial discussions on vortex and turbulence. The research is partly supported by the Visiting Scholar Scholarship of the China Scholarship Council (Grant No. 201808320079. The authors are grateful to Texas Advanced Computational Center (TACC) for providing computation hours. This work is accomplished by using code DNSUTA developed by Dr. Chaoqun Liu at the University of Texas at Arlington.

**References**

[1] Batchelor G. Introduction to Fluid Mechanics [M]. Cambridge: Cambridge University Press, 2000.
[2] Tennekes, H., Lumley, J. L. A First Course in Turbulence [M]. Massachusetts: MIT Press, 1972.
[3] Feynman R.F. The process. In: Gorter CJ, editor. Progress in low temperature physics, vol. 1 [M]. Amsterdam: North Holland Publishing Co. 1955.
[4] Liu C., Yan Y., and Lu P. Physics of turbulence generation and sustenance in a boundary layer [J]. *Computers & Fluids*, 2014: 102, 353-384.
[5] Liu C., Wang Y., Yang Y., et al. New omega vortex identification method [J]. *Science China Physics, Mechanics and Astronomy*, 2016, 59(8): 684711.
[6] Liu J., Wang Y., Gao Y., Liu C., Galilean invariance of Omega vortex identification method [J]. *Journal of Hydrodynamics,* 2019, 31(2): 249-255.
[7] Liu J., Gao Y., Wang Y., Liu C. Objective Omega vortex identification method [J]. *Journal of Hydrodynamics*, 2019, https://doi.org/10.1007/s42241-019-0028-y.
[8] Zhang Y., Qiu X., Chen P., et al. A selected review of vortex identification methods with applications [J]. *Journal of Hydrodynamics*, 2018, 30(5): 767-779.
[9] Zhang Y., Liu K., Li J., et al. Analysis of the vortices in the inner flow of reversible pump turbine with the new omega vortex identification method [J]. *Journal of Hydrodynamics*, 2018, 30(3): 463-469.
[10] Liu C., Gao Y., Tian S., and Dong X. Rortex—A new vortex vector definition and vorticity tensor and vector decompositions [J]. *Physics of Fluids*, 2018, 30: 035103.
[11] Gao Y., Liu C. Rortex and comparison with eigenvalue-based vortex identification criteria [J]. *Physics of Fluids*, 2018, 30: 085107.
[12] Wang Y., Gao Y., Liu J., Liu C. Explicit formula for the Liutex vector and physical meaning of vorticity based on the Liutex-Shear decomposition [J]. *Journal of Hydrodynamics*, 2019, https://doi.org/10.1007/s42241-019-0032-2.
[13] Liu C., Gao Y., Dong X. et al. Third generation of vortex identification methods: Omega and Liutex/Rortex based systems [J]. *Journal of Hydrodynamics*, 2019, 31(2): 205-223.